\documentclass[doublecol]{epl2} 
\usepackage{amsmath}
\usepackage{amssymb}
\usepackage{float}
\usepackage{color}
\usepackage{dcolumn}% Align table columns on decimal point
\usepackage{bm}% bold math
\usepackage{xcolor}
\usepackage{multirow}
\usepackage{adjustbox}
\usepackage{longtable}
\usepackage{comment}
\usepackage[normalem]{ulem}
\usepackage{array}
\usepackage[colorlinks,citecolor=blue,linkcolor=red,anchorcolor=blue,filecolor=blue,urlcolor=blue]{hyperref}

\title{Relativistic energy density functional from momentum space to coordinate space within a coherent density fluctuation model}
\shorttitle{Relativistic energy density functional at local density} %Insert here a short version of the title if it exceeds 70 characters

\author{Praveen K. Yadav\inst{1} \and Raj Kumar\inst{1} \and M. Bhuyan\inst{2}}
\shortauthor{Yadav, Kumar and Bhuyan}

\institute{                    
\inst{1} Department of Physics and Materials Science, Thapar Institute of Engineering and Technology, Patiala-147004, Punjab, India \\
\inst{2} Center for Theoretical and Computational Physics, Department of Physics, Faculty of Science, Universiti Malaya, Kuala Lumpur-50603, Malaysia
}

\abstract{\noindent
In this theoretical study, we have derived a simplified analytical expression for the binding energy per nucleon as a function of density and isospin asymmetry within the relativistic mean-field model. We have generated a new parameterization for the density-dependent DD-ME2 parameter set using the Relativistic-Hartree-Bogoliubov approach. Moreover, this work attempts to revisit the prior polynomial fitting in [Phy. Rev. C {\bf 103}, 024305 (2021)] for the non-linear NL3 force parameter to provide a simplified set of equations for the energy density functional which is used for calculating the surface properties of finite nuclei. The current study improves the existing fitting procedure by effectively proposing a simpler model that provides comparably precise results while lowering the computational expense. To study the surface properties of finite nuclei with these parameterizations, we have adopted the coherent density fluctuation model, which effectively translates the quantities of nuclear matter from momentum space to coordinate space at local density. The isospin properties, such as symmetry energy and its surface and volume components, slope parameter, finite nuclear incompressibility, and surface incompressibility for even-even nuclei, are calculated for different mass regions. Moreover, we have studied the effect of density, weight function, and choice of relativistic force parameters on the surface properties. The consequence of this work will help to determine the properties of nuclei along the nuclear landscape and can facilitate an improved understanding of the island of stability, heavy-ion collision, and nucleosynthesis, among others.}
\begin{document}
\maketitle
%%%%%%%%%%%%%%%%%%%%%%%%%%%
\section{Introduction}
The theoretical determination of infinite nuclear matter and finite nuclear properties has been at the forefront of research in nuclear spectroscopy and nuclear physics of intermediate energies \cite{gaid21,ivan18}. This presupposes a direct link between the density-dependent nuclear symmetry energy (NSE) and the isospin asymmetry for finite and infinite nuclear systems. In particular, the relevance of the NSE as a fundamental quantity stretches from the ground state structure of exotic nuclei to the physics of astrophysical systems such as neutron stars \cite{li19,zhan19}. Moreover, recent advances in the production of exotic beams \cite{decr91,reit20} have helped to understand the nuclear asymmetry of stable nuclei. Usually, an increase in the neutron-proton asymmetry marks the corresponding increase in the nuclear energy, leading to a variation in the density-dependent NSE and, thus, opening the avenue to extrapolate to the limit of neutron matter and evaluate the neutron star properties. In this vein, changes in the nuclear structure can influence the reaction dynamics of a system \cite{dani03}. Despite these variations, the constraints on NSE and other quantities of nuclear matter \cite{li13,dutr14} should not be compromised. This challenge could be addressed by a workable approach that can be employed from sub-saturation to supra-saturation densities.
%%%

When discussing the finite nucleus, most nuclear models properly fit in and/or near the $\beta-$ stable region and are extrapolated to exotic regions of the nuclear landscape. In general, traditional observables such as binding energy, separation energy, and related quantities are not particularly suitable parameters to precisely justify the existence of magicity or shell and/or sub-shell closure near the drip line region due to high isospin asymmetry \cite{otsu20,mich20}. To explain these nuclei, we need a particular observable that is connected to the $n-p$ asymmetry, that is, the symmetry energy and its coefficients. Thus, we require the inclusion of such properties in the study of finite nuclei. However, it is not easy to translate the infinite nuclear matter quantities to their corresponding finite nuclear quantities at local density. To tackle this problem, we need an approximated formalism to translate nuclear matter quantities from momentum space to coordinate space. Among several existing approaches \cite{myer66,dutr12,vret03}, the Br\"ueckner energy density functional (Br\"{u}ckner-EDF) \cite{brue68,brue69} within the coherent density fluctuation model (CDFM) \cite{bhuy18,prav22,prav23,gaid11} is notable for its successful treatment of surface properties by precisely translating the quantities from momentum space to coordinate space.
%%%%%%%

The CDFM approach has a wide range of applications in the structural study of finite nuclei near and far from the drip line, providing theoretical support for experimental data and even predicting new shell and sub-shell closures, providing a viable region of interest for experimental facilities \cite{bhuy18,prav22,gaid11}. Recently, the CDFM formalism has been successfully applied in calculating the incompressibility of finite nuclei using the giant monopole resonance (GMR) compression modes in nuclei \cite{gaid23}. It should be noted that most of the recent studies on CDFM involved the use of non-relativistic Br\"uckner's prescription \cite{gaid11,anto16}. Recently, relativistic inputs were employed within the CDFM formalism through the conventional Br\"uckner's prescription \cite{bhuy18,prav22}. However, the Br\"uckner's prescription is usually outfaced with the Coester-Band problem \cite{coes70,broc90} (failure to reproduce the precise binding energy per nucleon $B.E./A$ and empirical saturation density $\rho_0$), leading to inaccurate predictions of kinks in the symmetry energy which is used to signify the possible existence of shell closure. This Coester-band problem in heavy systems such as in the $Pb-$ isotopic chain is critically studied in Ref. \cite{patt22}. Therefore, one of our collaborators has recently proposed a relativistic energy density functional (relativistic-EDF) \cite{kuma21} based on the effective field theory motivated relativistic mean-field model. Interestingly, the relativistic-EDF framework has proved to be a formidable tool to confirm the presence of shell closures at the well-known neutron magic number $N$ = 126 and predicted notable kinks at $N$ = 172 and 184 \cite{patt22} while being adept for astrophysical applications, for example, neutron star systems \cite{kuma21a}.
%%%%%%%

The relativistic-EDF in Ref. \cite{kuma21}, used to parameterize with 24 ad-hoc parameters, however, yields marginally overestimated results in the case of finite nuclei. The work uses polynomial fitting to minimize the error while simultaneously linearly increasing the number of terms. In the present work, we revisit the previous fitting procedure \cite{kuma21} to present a new parameterization by adopting statistical techniques, especially principal component analysis (PCA) \cite{Pear01,abdi10}. We generate a new fitting parameter at local density using the Relativistic-Hartree-Bogoliubov (RHB) approach for the density-dependent DD-ME2 parameter set. It is noted that the DD-ME2 parameter set successfully provides a quantitative description of the properties of finite nuclei and infinite nuclear matter for a wide range of densities \cite{dutr14}. However, the use of density-dependent parameters is computationally more challenging, especially in the calculation of a random phase approximation as compared to the non-linear parametrizations \cite{lala09,nikv02}. It is also imperative that the nuclear matter quantities agree with their respective constraint range \cite{dutr14,bhar18,weic14}. In the present analysis, we have considered known $even-even$ nuclei, namely $^{16}$O, $^{40}$Ca, $^{48}$Ca, $^{56}$Ni, $^{90}$Zr, $^{116}$Sn, and $^{208}$Pb nuclei, to examine the surface properties at local density. These nuclei provide an opportunity to examine the $n-p$ asymmetry and confirm their influence on the properties of nuclei lying close to the $\beta$-stability line. This work has the potential to foster various studies that demand an accurate determination of the symmetry energy and its coefficients for the drip-line region of the nuclear chart and also the physics of neutron stars, the nucleosynthesis process, heavy-ion collision, and the island of stability of exotic nuclei.
%%%%%

%%%%%%%%%%%%%%%%%%%%%%%%%%%%%%%%%%%%%
\section{Theoretical Formalism}
\label{theory}
\noindent
The relativistic mean-field (RMF) formalism constitutes a microscopic approach to solving the many-body problem through the interacting meson fields. The RMF formalism is widely used in studying finite nuclei and infinite nuclear matter, including neutron star systems. It can be classified as the relativistic interpretation of Hartree-Fock Bogoliubov's theory based on the medium effect. In-depth details of the RMF formalisms and their parameterization are given in Refs. \cite{ring96,bogu77,lala09,nikv02}. A characteristic RMF Lagrangian density established subsequently with several changes to the original Walecka Lagrangian has the form (see Refs.\cite{ring96,bhuy18,prav22,bogu77,lala09,nikv02}):
\begin{eqnarray}
	\label{eqt:Lag}
	{\cal L}&=&\overline{\psi}\{i\gamma^{\mu}\partial_{\mu}-M\}\psi +{\frac12}\partial^{\mu}\sigma
	\partial_{\mu}\sigma \nonumber \\
	&& -{\frac12}m_{\sigma}^{2}\sigma^{2}-{\frac13}g_{2}\sigma^{3} -{\frac14}g_{3}\sigma^{4}
	-g_{s}\overline{\psi}\psi\sigma \nonumber \\
	&& -{\frac14}\Omega^{\mu\nu}\Omega_{\mu\nu}+{\frac12}m_{w}^{2}\omega^{\mu}\omega_{\mu}
	-g_{w}\overline\psi\gamma^{\mu}\psi\omega_{\mu} \nonumber \\
	&&-{\frac14}\vec{B}^{\mu\nu}.\vec{B}_{\mu\nu}+\frac{1}{2}m_{\rho}^2
	\vec{\rho}^{\mu}.\vec{\rho}_{\mu} -g_{\rho}\overline{\psi}\gamma^{\mu}
	\vec{\tau}\psi\cdot\vec{\rho}^{\mu}\nonumber \\
	&&-{\frac14}F^{\mu\nu}F_{\mu\nu}-e\overline{\psi} \gamma^{\mu}
	\frac{\left(1-\tau_{3}\right)}{2}\psi A_{\mu}.
	\label{lag}
\end{eqnarray}
with vector field tensors $F^{\mu\nu} = \partial_{\mu} A_{\nu} - \partial_{\nu}A_{\mu}$, $\Omega_{\mu\nu} = \Omega_{\mu\nu} \partial_{\mu} \omega_{\nu} - \partial_{\nu} \omega_{\mu}$ and $\vec{B}^{\mu\nu} = \partial_{\mu} \vec{\rho}_{\nu} - \partial_{\nu} \vec{\rho}_{\mu}$. 
%\begin{eqnarray}
%	F^{\mu\nu} = \partial_{\mu} A_{\nu} - \partial_{\nu} A_{\mu} \nonumber \\
%	\Omega_{\mu\nu} = \partial_{\mu} \omega_{\nu} - \partial_{\nu} \omega_{\mu} \nonumber \\
%	\vec{B}^{\mu\nu} = \partial_{\mu} \vec{\rho}_{\nu} - \partial_{\nu} \vec{\rho}_{\mu}.
%\end{eqnarray}
The term $g_{\sigma}$, $g_{\rho}$ and $g_{\omega}$ in Eq. (\ref{eqt:Lag}) refers to the coupling constant of $\sigma$, $\rho$ and $\omega$ meson respectively. 
From the above Lagrangian, one can acquire the field equations for the nucleons and the mesons by expanding the upper and lower components of the Dirac spinors and the boson fields \cite{ring96,lala09,nikv02}. The RMF formalism permits the density dependence of the meson-nucleon coupling, as described in Refs. \cite{lala05}. This coupling is parameterized in the phenomenological approach to the nucleon fields as:
\begin{eqnarray}
	\centering 
	\label{eqn:gp_ddme2}
	g_{i}(\rho)=g_{i}(\rho_{sat})f_{i} (x)\vert_{i=\sigma,\omega},
\end{eqnarray}
where
\begin{eqnarray}
	\centering 
	\label{eqn:fix_ddme2}
	f_{i}(x)=a_{i}\dfrac{1+b_{i}(x+d_{i})^{2}}{1+c_{i}(x+d_{i})^2},
\end{eqnarray}
and
\begin{eqnarray}
	\centering 
	\label{eqn: grho_ddme2}
	g_{\rho}=g_{\rho}(\rho_{sat})e^{a_{\rho}(x-1)}.
\end{eqnarray}	    
%%%    
A detailed description of relativistic Hartree-Fock Bogoliubov formalism, including its solutions, is given in Refs. \cite{lala05,nikv02,bhuy18}. In the present work, two different pairing approaches are taken into account to examine the effect of pairing and model dependencies on the quantities of nuclear matter at local density. The constant-gap BCS technique with NL3 and the Bogoliubov transformation with the DD-ME2 parameter are considered while accounting for pairing correlations to characterize the nuclear bulk properties of open-shell nuclei. A more detailed review of relativistic parameterization and formalism can be found in Refs. \cite{ring96,bhuy18,bogu77,lala09,nikv02}.  %%%%%%%%%%%%%%%%%%%%%%%%%%%%%%%%%%%%%%%%%%%%%%%%%%
\subsection{Relativistic energy density parameterization at local density} \label{ssec:fitting}
The recent work \cite{kuma21}, which dealt with the polynomial fitting of binding energy per nucleon mainly focused on lowering the mean deviation of the fitted nuclear matter data with respect to the calculated values. This results in a linearly increasing number of terms of both a$_{i}$ and b$_{i}$ as depicted in Eq. (6) of Ref. \cite{kuma21}. In the present work, we employ the PCA \cite{Pear01,abdi10,dese95,saye15} to have non-linear parametrization while using fewer predictors, which drastically improves the calculation speed and provides comparably precise results. PCA is a type of unsupervised learning technique that analyzes how a set of variables are related to each other using the correlation matrix. It is also called a general factor analysis, where regression finds the best-fitting line. It is the most common tool for exploratory data analysis and machine learning for predictive models. The main aim of PCA is to discover a new set of variables that are smaller than the original set of variables but contain most of the information in the sample and are useful for the regression and classification of data. More details regarding the utilization and implementation of PCA can be found in Ref. \cite{Pear01,abdi10,dese95}. 
%%%%%%%%%

It is interesting to note that the physics models typically have a low variance and higher bias, while the artificially derived models typically have a high variance and a low bias \cite{dris19}. When dealing with parameter minimization, it is recommended to follow the principle of Occam's razor \cite{cull93,blum87}, which states that for given two models with the same generalized error, one should choose the simpler model over the complex counterpart. The performance of the modelling algorithm degrades with an increase in the number of predictors. This is also called the `Curse of Dimensionality' \cite{nisb18}. An overly complex model increases the difficulty of dealing with experimental data, as it places an additional demand for measuring predictors, which wastes resources and could result in undetected errors in the results. Moreover, a model involving fewer terms can be easily reproducible, while a complex model may require a large amount of calibration and specific software to estimate the model effectively. Thus, one can generate overall lower prediction error models by carefully incorporating physics-based constraints into artificial models or vice-versa.
%%%%%%%%%%%%%%%%%%%%%%%%%%%%%%%%%%%%%%

\begin{figure}[htbp]
	\includegraphics[scale=0.36
	]{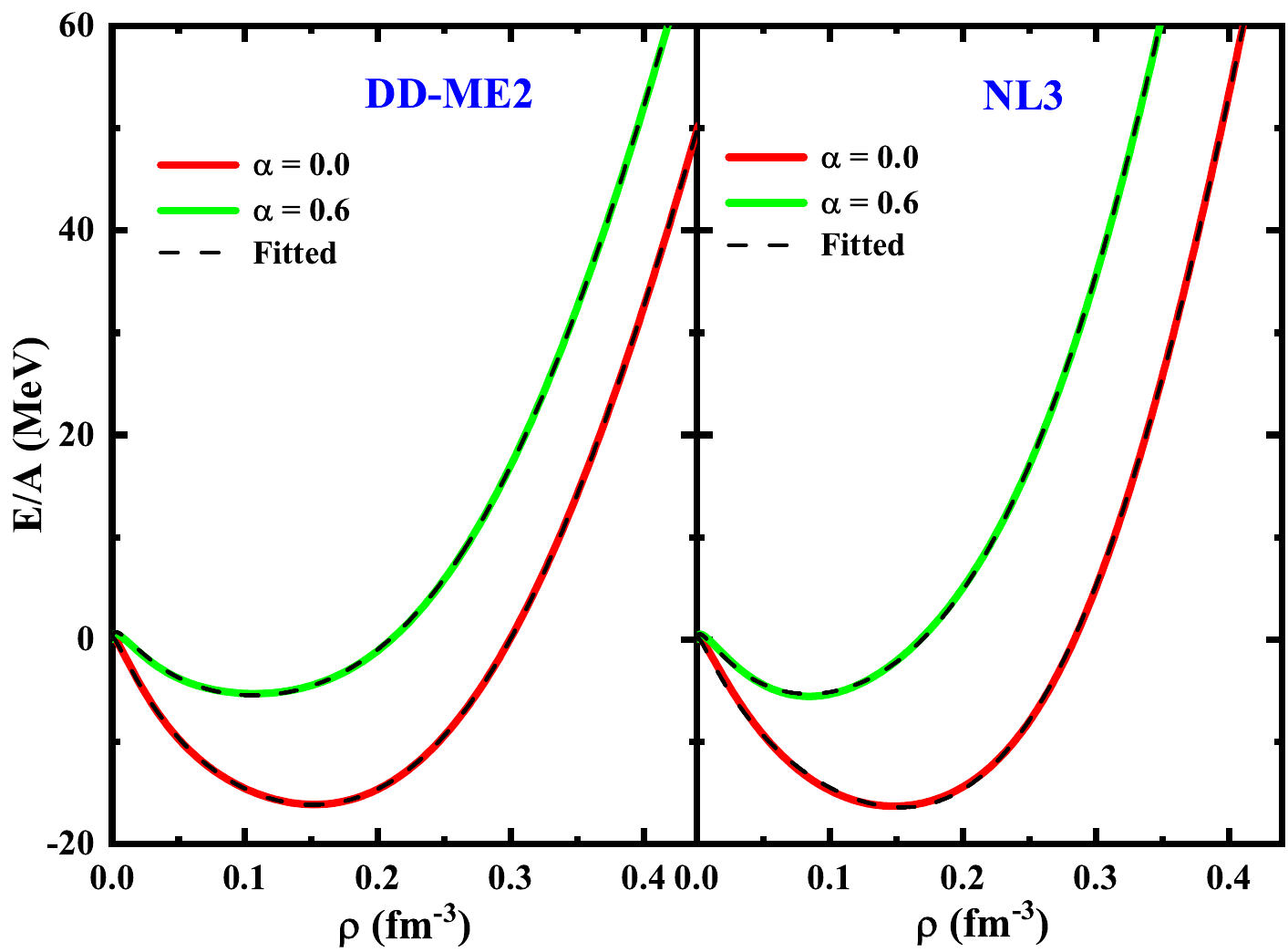}
	\caption{The nuclear matter $ E/A $ as a function of total number density ($\rho_{n}$+$\rho_{p}$)  of DD-ME2 and NL3 parameter sets for different asymmetry $ \alpha=\dfrac{\rho_{n}-\rho_{p}}{\rho_{n}+\rho_{p}} $ based on new 8 term fitting.}
	\label{fig_fit}
	\vspace{-0.4cm}
\end{figure}

In this study, following the PCA detailed in Refs. \cite{abdi10,dese95,saye15}, we use correlation matrices and correlation coefficients, to find the interdependence of ad-hoc terms and minimize the fitting expression of the EDF.  For studying the deviation of the minimized fitting expression, we adopt the Levenberg-Marquardt iteration algorithm as detailed in Ref. \cite{marq63,more78}. However, considering the nucleus as a collection of tiny spherical pieces described by a local density functional $\rho_{0}(x)=3A/4\pi x^{3}$, the expression for the fitted EDF for NL3 and DD-ME2 force parameter sets within the RMF formalism is stated as:
%%%%
\begin{align}
	\label{eqn:fit}
	\varepsilon (\rho)=& C_{k}\rho_{0}^{2/3}(x)+b_{1}\rho_{0}(x)+b_{2}\rho_{0}^{5/3}(x)+b_{3}\rho_{0}^{8/3}(x)\nonumber\\
	&
	+b_{4}\rho_{0}^{10/3}(x)+b_{5}\rho_{0}^{4}(x) \nonumber\\
	&
	+\alpha^{2}\Bigl(a_{1}\rho_{0}^{2/3}(x)+a_{2}\rho_{0}^{7/3}(x)+a_{3}\rho_{0}^{8/3}(x)\Bigl).
\end{align}
%%%
Here, the first entity $C_{k}$ implies kinetic energy term from the Thomas-Fermi approximation \cite{hohe64,brue68,brue69} having unit MeVfm$^{-1}$ with  $C_{k}=37.53[(1+\alpha)^\frac{5}{3}+(1-\alpha)^\frac{5}{3}]$ and $\alpha$ refer to the neutron-proton asymmetry. The coefficients $b_{i}$ correspond to the potential energy terms of the fitting equation, where $i$ varies from 1 to 5. These terms describe the interaction between nucleons mediated by meson fields, such as the scalar $\sigma-$, the vector $\omega-$, and the isovector $\rho-$ mesons. They also include the non-linear terms that simulate the three-body effects in the nuclear potential. These terms have different powers of the nuclear density, ranging from 1 to 4, and they determine the shape of the energy density curve as a function of nuclear density (Fig. \ref{fig_fit}). Furthermore, the coefficients $a_{i}$ correspond to the symmetry energy part of the potential energy in the fitting equation, where $i$ varies from 1 to 3. These $a_{i}$ terms describe the dependence of $E/A$ on $\alpha$, which measures the deviation from symmetric nuclear matter ($\alpha$ = 0) to pure neutron matter ($\alpha$ = 1). Furthermore $a_{i}$ terms have different powers of the nuclear density, ranging from 2/3 to 8/3, and are crucial in determining the magnitude and slope of the symmetry energy.
%%%%%%
%%%%%%%%%%%%%%%%%%%%%%%%%%%%%%%%%%%%%%%%%%%%%%%%%%%%%%%%
%%%%%%%%%%%%%%%%%%%%%%%%%%%%%%%%%%%%%%%%%%%%%%%%%%%%%%%%

The fitting procedure took 12 iterations for NL3 and 13 iterations for the DD-ME2 parameter set. The root-mean-square deviation (RMSD) is calculated using the formula $RMSD = \sqrt{\dfrac{1}{N}\sum_{i=1}^{N}[(E/A)_{i,Fitted}-(E/A)_{i,RMF}}]$. In this eight-term fitting work, the RMSD for the binding energy is 0.07704 MeV for DD-ME2 and 0.19146 MeV for the NL3 parameter set, providing the best fit. In the present work, individually fitting the energy density functional of DD-ME2 and NL3 parameter sets is comparatively less resource extensive. This implies that finding an appropriate fitting procedure that can simultaneously define both parameter sets turns out to be time-consuming and resource-intensive. This is due to the difference in the curvature of the graph corresponding to these parameter sets for different neutron-proton asymmetry. To find the optimal fitting procedure, we use a correlation matrix with correlation coefficients that describe the one-to-one relationship between the various predictors. We simultaneously reduce the strongly correlated parameters while adding newer terms that can effectively describe the nuclear matter properties. In every subsequent step, we performed the calculations and minimized the RMSD. We found that eight terms yield the highest accuracy and reproduce reliable results for various properties of nuclear matter.
%%%%%%%%%%%%%%%%%%%%%%%%%%%%%%%%%%%%%%%%%%%%
%%%%%%%%%%%%%%%%%%%%%%%%%%%%%%%%%%%%%%%%%%%%%%%%%%%%%%%%%%%%%%
\begin{table}[htbp]
	\centering
	\caption{The coefficients obtained from the optimal 8-term fitting procedure for the nuclear binding energy per particle $E/A$ as a function of nuclear density $\rho_{0}(x)$ and neutron-proton asymmetry parameter $\alpha$. The values of the coefficient with their units are listed for DD-ME2 and NL3 parameter sets.}
	\renewcommand{\tabcolsep}{0.35cm}
	\renewcommand{\arraystretch}{1.3}
	\begin{tabular}{|c|r|r|r|}
		\hline \hline
		& \multicolumn{1}{|c|}{DD-ME2} & \multicolumn{1}{|c|}{NL3} & \multicolumn{1}{|c|}{Unit} \\
		\hline
		b1 & -627.40397 & -631.22898 & MeV \\
		\hline
		b2 & 2032.92832 & 2177.46092 & MeVfm$^2$ \\
		\hline
		b3 & -9038.76463 & -11541.44105 & MeVfm$^5$ \\
		\hline
		b4 & 19143.35052 & 26104.23289 & MeVfm$^7$ \\
		\hline
		b5 & -12352.13859 & -17045.86031 & MeVfm$^9$ \\
		\hline
		a1 & 80.43433 & 49.40867 & MeVfm$^{-1}$ \\
		\hline
		a2 & -433.81712 & 2741.5454 & MeVfm$^4$ \\
		\hline
		a3 & 446.43825 & -3019.76641 & MeVfm$^5$ \\
		\hline \hline
	\end{tabular}%
	\label{tab:fit_para}
	\vspace{-0.4cm}
\end{table}%
%%%%%%%%%%%%%%%%%%%%

The nuclear matter parameters $ K^{NM} $, $ S^{NM} $, $ L^{NM}_{sym} $ and $ K^{NM}_{sym} $ can be derived from standard relations in Refs. \cite{bhuy13,bhar18,weic14} by using Eq. (\ref{eqn:fit}), 
\begin{align}
	\label{knm_sol}
	K^{NM}(x) = & -150.12\rho_{0}^{2/3}(x) + 10b_{2}\rho_{0}^{5/3}(x) + 40b_{3}\rho_{0}^{8/3}(x)  \nonumber\\
	& + 70b_{4}\rho_{0}^{10/3}(x) + 108 b_{5}\rho_{0}^{4}(x),
\end{align}
%%%%
\begin{align}
	%	\centering 
	\label{snm_sol}
	S^{NM}(x) = & 41.7\rho_{0}^{2/3}(x) + a_{1}\rho_{0}^{2/3}(x) \nonumber\\
	& + a_{2}\rho_{0}^{7/3}(x) + a_{3}\rho_{0}^{8/3}(x),
\end{align}	
%%%
\begin{align}
	%	\centering 
	\label{lnm_sol}
	L^{NM}_{sym}(x) = & 83.4\rho_{0}^{2/3}(x) + 2a_{1}\rho_{0}^{2/3}(x) \nonumber\\
	& + 7a_{2}\rho_{0}^{7/3}(x) + 8a_{3}\rho_{0}^{8/3}(x),
\end{align}	
%%%
\begin{align}
	%	\centering 
	\label{knm_sym_sol}
	K^{NM}_{sym}(x) = & -83.4\rho_{0}^{2/3}(x) -2a_{1}\rho_{0}^{2/3}(x) \nonumber\\
	&  + 28a_{2}\rho_{0}^{7/3}(x) + 40a_{3}\rho_{0}^{8/3}(x).
\end{align}
%%%%
To note, we incorporate CDFM formalism, discussed in the subsequent subsection, to translate the infinite nuclear matter quantities in the realm of finite nuclei. 
%%%%%%%
\subsection{Coherent density fluctuation model} \label{ssec:cdfm}
The coherent density fluctuation model (CDFM) is a natural extension of Fermi-gas model which is developed by Antonov {\it et al.} \cite{anto78,anto80, anto79,anto82,gaid11}. It is based on the $ \delta $-function limit of the generator coordinate method \cite{grif57}. According to the CDFM, nuclear matter exhibits density fluctuations around an average distribution, while maintaining both spherical symmetry and uniformity. The calculation assumes that the nuclear matter is composed of tiny spheres of nuclear matter, which are referred to as ``fluctons" with local density function $\rho_{o}(x)=\dfrac{3A}{4\pi x^{3}}$ \cite{gaid11,anto79,anto82,bhuy18}. The weight function $\vert \mathcal{F}(x) \vert^{2}$ of a nucleus is calculated as:
\begin{eqnarray}
	\centering 
	\label{eqn:21}
	\vert \mathcal{F}(x) \vert^{2} = -\frac{1}{\rho_{o}(x)}\frac{d\rho(r)}{dr} \bigg|_{r=x},
\end{eqnarray}
with normalization $\int_{0}^{\infty}dx \vert \mathcal{F}(x) \vert^{2} = 1$. The weight function serves as the crucial bridge between the infinite nuclear matter quantities described in momentum space and the corresponding finite nuclear matter quantities described in coordinate space. For detailed analytical derivation for obtaining the density-dependent weight function, one can refer to Refs. \cite{anto82,anto80,anto78,anto94}.
%%%%%%%%%%%%%%%%%%%%%%%%%%%%%%%%%%%%%%%%%%%%%%%%%%%%%%%%%%

In the CDFM formalism, the finite nuclear incompressibility $K^{A}$, the effective symmetry energy $ S $, the widely used slope parameter $L_{sym}^{A}$ and surface incompressibility $K_{sym}^{A}$ is calculated by weighting the corresponding quantities of the infinite nuclear matter as \cite{anto94, gaid11, sarr07,bhuy18}:
\begin{eqnarray}
	\label{eqn:Kf}
	K^{A} = \int_{0}^{\infty}dx \vert\mathcal{F}(x)\vert^{2}K^{NM}(x).
\end{eqnarray}
%%%
\begin{eqnarray}
	\label{eqn:Sf}
	S = \int_{0}^{\infty}dx \vert\mathcal{F}(x)\vert^{2}S^{NM}(x).
\end{eqnarray}
%%%
\begin{eqnarray}
	\label{eqn:Lf}
	L^{A}_{sym} = \int_{0}^{\infty}dx \vert\mathcal{F}(x)\vert^{2}L^{NM}_{sym}(x).
\end{eqnarray}
%%%
\begin{eqnarray}
	\label{eqn:Kaf}
	K^{A}_{sym} = \int_{0}^{\infty}dx \vert\mathcal{F}(x)\vert^{2}K^{NM}_{sym}(x).
\end{eqnarray}

%%%%%%%%%%%%%%%%%%%%%%%%%%%%%%
Following the Danielwicz's prescription, the volume and surface composition of the symmetry energy can be expressed as \cite{dani03, dani06,dani09}:
\begin{eqnarray}
	\centering 
	\label{eqn:Sv}
	S_{V}=S\bigg(1+\dfrac{1}{\kappa A^{1/3}}\bigg),
\end{eqnarray}
and    
\begin{eqnarray}
	\centering 
	\label{eqn:Ss}
	S_{S}=\dfrac{S}{\kappa}\bigg(1+\dfrac{1}{\kappa A^{1/3}}\bigg),
\end{eqnarray}
respectively. Here, the term $\kappa\equiv S_{V}/S_{S}$ is the ratio of volume symmetry energy to that of the surface symmetry energy. More details related to the CDFM formalism can be found in Ref. \cite{gaid11,anto16,gaid21,prav22,prav23,bhuy18}. 
%%%%%%%%

\section{Calculations and Results}
\label{Results} \noindent
The present work focuses on the derivation of a new optimized parameterization of the relativistic energy density functional for effectively studying nuclear matter quantities. This work provides new parameterization based on the relativistic-EDF of the DD-ME2 set and revisits the recent fitting procedure based on the NL3 set. It is crucial to note that prior calculations with Br\"{u}ckner-EDF can reveal inconsistent results in a few nuclei due to the Coester-Band problem \cite{coes70,broc90}. This means that the saturation curves of nuclear matter related to the classical Br\"{u}ckner fails to accurately recreate the empirical saturation point of nuclear matter that is $E/A \approx$ -16 MeV occurs near about $\rho \approx 0.2$ fm$^{-3}$ instead of $\rho \approx 0.15$ fm$^{-3}$.
%%%%

Here, it is important to understand the principal reason for the emphasis on resolving the Coester-band problem. The original work on the CDFM formalism is primarily applied using Brückner's prescription \cite{gaid11}. Even in the later incorporation of the RMF formalism within the CDFM by taking the RMF densities \cite{bhuy18,prav22}; still, the original Br\"ueckner's prescription, which was baked in the CDFM, was employed. The importance of a feature complete implementation of relativistic prescription within CDFM formalism is emphasized in one of our recent works (Ref. \cite{patt22}), which studies the surface properties of the Pb- isotopic chain and draws a thorough comparison on the application of relativistic-EDF to counter the Coester band problem plaguing the non-relativistic prescription. Using Brückner's functional, the calculated symmetry energy and its components show an incorrect peak at $N$ = 120. However, the relativistic-EDF successfully resolves the issue to give an accurate peak at $N$ = 126. The direct advantage of relativistic-EDF is that the non-linear terms of the RMF Lagrangian can simulate the nuclear potential's three-body effect and fit to the Coester-Band region \cite{pipe01}. 
%%%%
%%%%%%%%%%%%%%%%%%%%%%%%%%%%%%%%%%%%%%%%%%%%%%%%%%%%%%%% 
\begin{figure}[htpb]
	\centering
	\includegraphics[width=8.3cm,height=8cm]{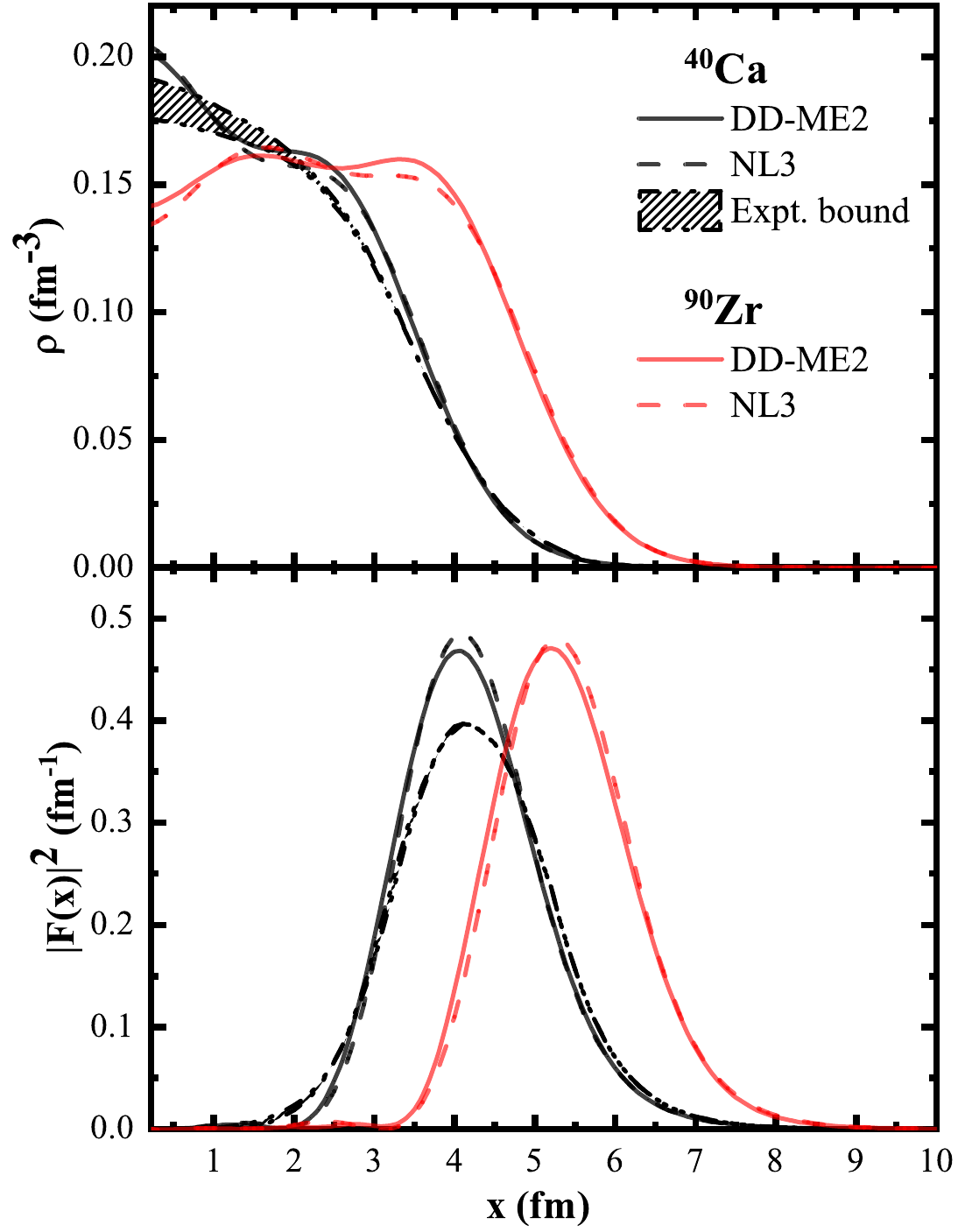}
	\caption{Total density distribution (upper panel) and corresponding weight function (lower panel) as a function of nuclear distance for $^{40}$Ca and $^{90}$Zr nuclei as representative case using density-dependent DD-ME2, and non-linear NL3 parameter sets compared with available experimental bound data.}
	\label{fig_density}
	\vspace{-0.4cm}
\end{figure}
%%%%%%%%%%%%%%%%%%%%%%%%%%%%%%%%%%%%%%%%%%	
Fig. \ref{fig_fit} shows the graph of fitted nuclear matter $E/A$ as a function of the nuclear density based on the DD-ME2 and NL3 parameters for varying neutron-proton asymmetry $\alpha$. Here, we fit the terms of $E/A$ based on non-linear polynomial fitting, where the first term corresponds to the kinetic energy obtained from the Thomas-Fermi approximation, and the subsequent terms are part of the potential energy. The values and units of the coefficients corresponding to the potential energy are given in Table \ref{tab:fit_para}.
%%%%%%%%%%%%%%%%%%%%%%%%%%
\begin{table}[htbp]
	\centering
	\caption{The finite nuclear incompressibility $K^{A}$, symmetry energy $S$, slope parameter $L^{A}_{sym}$, surface incompressibility $K^A_{sym}$, and surface $S_{S}$ and volume $S_{V}$ components of symmetry energy in MeV, using the relativistic energy density functional with density-dependent DD-ME2 and non-linear NL3 force parameters.}
	\renewcommand{\tabcolsep}{0.095cm}
	\renewcommand{\arraystretch}{1.3}
	\begin{tabular}{cccccccc}
		\hline \hline
		DD-ME2 & $^{16}$O & $^{40}$Ca & $^{48}$Ca & $^{56}$Ni & $^{90}$Zr & $^{116}$Sn & $^{208}$Pb \\
		\hline
		$K^{A}$ & 291.0 & 245.7 & 267.6 & 327.3 & 256.9 & 254.9 & 208.1 \\
		$S^{A}$ & 22.7 & 22.8 & 23.4 & 25.4 & 24.1 & 24.2 & 23.4 \\
		$L^{A}_{sym}$ & 38.7 & 39.1 & 39.8 & 42.7 & 40.9 & 41.2 & 40.2 \\
		$K^{A}_{sym}$ & -65.7 & -69.1 & -71.7 & -77.9 & -74.5 & -75.1 & -72.4 \\
		$S_{V}$ &  28.1 & 26.7 & 27.5 & 29.9 & 27.4 & 27.2 & 25.5 \\
		$S_{S}$ & 16.7 & 15.2 & 17.1 & 19.9 & 16.5 & 15.9 & 13.4 \\
		\hline \hline
		NL3   & $^{16}$O & $^{40}$Ca & $^{48}$Ca & $^{56}$Ni & $^{90}$Zr & $^{116}$Sn & $^{208}$Pb \\
		\hline
		$K^{A}$ & 319.2 & 267.3 & 348.3 & 382.7 & 282.3 & 252.8 & 228.6 \\
		$S^{A}$ & 27.2 & 26.7 & 30.3 & 32.1 & 28.8 & 27.8 & 27.6 \\
		$L^{A}_{sym}$ & 85.7 & 85.4 & 97.5 & 103.8 & 93.1 & 89.6 & 88.8 \\
		$K^{A}_{sym}$ & 27.0 & 45.3 & 41.4 & 43.3 & 54.1  & 56.7 & 60.9 \\
		$S_{V}$ & 33.6 & 31.5 & 35.9 & 37.9 & 33.1 & 31.4 & 30.2 \\
		$S_{S}$ & 19.9 & 19.2 & 23.9 & 26.2 & 21.5 & 19.5 & 15.9 \\
		\hline \hline
	\end{tabular}%
	\label{tab:result}
\end{table}%
%%%%%%%%%%%%%%%%%
Following the CDFM formalism, we have calculated the isospin-dependent surface quantities such as finite nuclear incompressibility $K^{A}$, symmetry energy $S$, slope parameter $L^{A}_{sym}$, surface incompressibility $K^A_{sym}$, and surface $S_{S}$ and volume $S_{V}$ components of symmetry energy for a few double closed-shell nuclei namely $^{16}$O, $^{40}$Ca, $^{48}$Ca, $^{56}$Ni, $^{90}$Zr, $^{116}$Sn and $^{208}$Pb. The CDFM effectively addresses the fluctuations in the values of nuclear matter properties near the surface of finite nuclei introduced by the density distribution through the weight function \cite{gaid11,bhuy18,prav22}.
%%%%%

In Fig. \ref{fig_density}, we present the density $\rho$ plot in the upper panel along with the corresponding weight function $\vert \mathcal{F} (x) \vert^{2}$ in the lower panel as a function of the nuclear distance $x$. Here, we have shown the nuclei $^{40}Ca$, and $^{90}Zr$ as representative cases.
From a careful inspection of the figure, one may notice that the weight function resembles a bell-shaped form with the maxima of the weight function at a given nuclear distance lying in the corresponding domain with a very low value of nuclear density. This location of the maxima of the weight function implies that the substantial part of the weight function lies in the surface region of the density; hence the name surface properties.

Following Eqs. (\ref{eqn:Kf}-\ref{eqn:Ss}), we next compute the surface properties, which are listed in Table \ref {tab:result}. From the table, one can easily observe the variation in the effective surface properties for given even-even nuclei, which is attributed to the density and weight function. In Refs. \cite{ston14,dutr14}, the value of $K_{0}$ has the range 230 $\pm$ 40 MeV. Specifically, the value of $K_{0}$ for NL3 is 271.53 MeV at saturation -16.24 MeV, whereas for DD-ME2  is 250.92 MeV at saturation -16.14 MeV \cite{dutr14}. Although the finite nuclei do not have a precise range of incompressibility of finite nuclei $K^{A}$, we still find that the range of $K^{A}$ is nearly consistent with $K_{0}$ for the case of DD-ME2 and NL3 force parameters.
Following Fig. \ref{fig_density} and Eq. (\ref{eqn:21}), we find that the density distribution has a peak at the center of the nucleus and decreases towards the surface, where it drops sharply to zero. The weight function has a peak value in the range that corresponds to the surface part of the density, where the density fluctuations are most pronounced. The slope of the density distribution affects the magnitude of the weight function in the central and surface regions, which in turn affects the magnitude of the calculated surface properties. The slope of the density distribution determines the surface thickness of the nucleus, which is the distance over which the density changes from the central value to the surface value. The surface thickness affects the weight function because it determines the range of possible local densities that can be found in the surface region. A steeper slope means a thinner surface with a narrower range of local densities, implying a higher probability of finding a certain local density, while a flatter slope means a thicker surface with a wider range of local densities, implying a lower probability of finding a certain local density. As the weight function is inherently the probability distribution of finding a coherent state with a given local density in the nucleus, therefore, we can infer that a thinner surface leads to a larger weight function and surface properties, while a thicker surface leads to a smaller weight function and surface properties.

On careful observation of the density and weight function, the slope of density given by the NL3 parameter is larger than that of DD-ME2, contributing to a larger value of the weight function. This large value of the weight function is reflected in the large value of the calculated surface properties corresponding to the NL3 parameter. Moreover, Eqs. (\ref{eqn:Kf})-(\ref{eqn:Kaf}) discuss the calculation of surface properties using the weight function and corresponding nuclear matter (NM) parameters (given in Eqs. (\ref{knm_sol})-(\ref{knm_sym_sol})). Thus, the surface properties depend on the weight function and the nuclear matter parameters. Since NL3 has a stiffer equation of state as compared to its counterpart (inferred from Fig. \ref{fig_fit} and Ref. \cite{type05}), it contributes to having larger values of the NM parameters. As the NL3 provides larger-valued NM parameters, these NM parameters multiplied with the corresponding larger value of the weight function effectively yield larger surface properties, as shown in Table \ref{tab:result}. Moreover, it is interesting to note the absence of distinct mass dependence from light to heavy mass nuclei (that is, from $^{16}$O to $^{208}$Pb) for the calculated quantities, which may be attributed to the structural effects \cite{bhuy18,prav22} that plays a crucial role in the distribution of density in a finite nucleus throughout the nuclear landscape.
%%%%%%%%%%%%%%%%%%%%%%%%%%%%%%%%%%%%%%%%%%%%%%%%%%%%%%

\section{Summary and conclusion}
\label{summary} \noindent
This work established a new optimized parameterization of the relativistic-EDF for the density-dependent DD-ME2 parameter set and revisits the previous fitting procedure for the widely used non-linear NL3 parameter set at local density. The motivation of this work is to provide a simplified expression of EDF and its derived surface properties, which helps to better understand the physics based on the parameters and also improve the computational cost. We have minimized the number of coefficient terms in relativistic energy density functional to the barest minimum of about one-third of the previous work \cite{kuma21}. These parameterizations are employed within the CDFM for estimating various nuclear matter properties, including nuclear symmetry energy, finite nuclear incompressibility, widely used slope parameters, and surface incompressibility for a few double closed-shell nuclei. We find that surface properties share a close relationship with density and its corresponding weight function. A decrease in surface thickness is associated with an increase in weight function and surface properties, whereas an increase in surface thickness is associated with a decrease in weight function and surface properties. This helps in having a precise determination of the symmetry energy and its coefficients which is essential in studying the island of stability, dipole polarizability, the physics of neutron stars, nucleosynthesis, and heavy-ion collisions. An in-depth investigation is being conducted shortly by utilizing the novel parameterization of relativistic energy density functional within CDFM formalism for nuclei in the different mass regions across the nuclear landscape.
%%%%%%%%%%%%%%%%%%%%%%%%%%%%

\acknowledgments
PY would like to thank J. T. Majekodunmi for his constructive effort in the drafting of the manuscript. This work has been supported by Science and Engineering Research Board (SERB) File No. CRG/2021/001229; FOSTECT Project Code.: FOSTECT.2019B.04; and FAPESP Project No. 2017/05660-0.

%%%%%%%%%%%%%%%%%%%%%%%%	%%%%%%%%%%%%%%%%%%%%%%%%	

\end{document}